\documentclass[12pt]{article}

\usepackage{epsfig,amsmath,amssymb,verbatim,mathrsfs}

\def\beq{\begin{eqnarray}}
\def\eeq{\end{eqnarray}}
\def\bea{\begin{eqnarray}}
\def\eea{\end{eqnarray}}

\def\zpri{Z'}

\def\tev{\, {\rm TeV}}
\def\gev{\, {\rm GeV}}
\def\mev{\, {\rm MeV}}
\def\kev{\, {\rm keV}}

\def\xfb{\, {\rm fb}}
\newcommand{\gsim}{\lower.7ex\hbox{$\;\stackrel{\textstyle>}{\sim}\;$}}
\newcommand{\lsim}{\lower.7ex\hbox{$\;\stackrel{\textstyle<}{\sim}\;$}}

\begin{document}

\setlength{\baselineskip}{0.2in}


\begin{titlepage}
\noindent
\begin{flushright}
CERN-PH-TH-2008-47 \\
MCTP-07-51  \\
\end{flushright}
\vspace{1cm}

\begin{center}
  \begin{Large}
    \begin{bf}
    How to Find a Hidden World \\ 
at the Large Hadron Collider
     \end{bf}
  \end{Large}
\end{center}
\vspace{0.2cm}
\begin{center}
\begin{large}
James D. Wells\\
\end{large}
  \vspace{0.3cm}
  \begin{it}
MCTP, University of Michigan, Ann Arbor, MI 48109 \\
CERN, Theory Division, CH-1211 Geneva 23, Switzerland

\vspace{0.1cm}
\end{it}

\end{center}


\begin{abstract}
I discuss how the Large Hadron Collider era should broaden our view of  particle physics research, and apply this thinking to the case of Hidden Worlds.  I focus on one of the simplest representative cases of a Hidden World, and detail the rich implications it has for LHC physics, including universal suppression of Higgs boson production, trans-TeV heavy Higgs boson signatures, heavy-to-light Higgs boson decays, weakly coupled exotic gauge bosons, and Higgs boson decays to four fermions via light exotic gauge bosons.  Some signatures may be accessible in the very early stages of collider operation, whereas others motivate a later high-lumonosity upgrade.

\end{abstract}

\vspace{1cm}

\noindent 
To be published as chapter in {\it Perspectives on LHC Physics}, edited by G.\ Kane and A.\ Pierce, World Scientific Publishing Co., 2008.


\end{titlepage}

\setcounter{page}{2}


\vfill\eject

\section*{Particle Physics in the LHC Era}

The annals of particle physics are replete with exhortations to solve the hierarchy problem, the flavor problem, the baryon asymmetry problem, the dark matter problem, the unification problem, etc.
Much of our  efforts go into constructing the simplest solution to one of these problems.  There is a premium on taut constructions narrowly tailored to solve our most precious problems.  

With the coming of the LHC era, electroweak symmetry breaking and naturalness become the central focus for at least the initial phase of running. Our community has had many ideas, the simplest being that a single scalar boson condenses to break the electroweak symmetry and simultaneously gives mass to all elementary particles. Although a logical possibility, few believe the Higgs boson alone is a viable option since it is so delicate to quantum corrections.  For thirty years the beyond-the-Standard Model  community has pursued various scenarios that support and protect the Higgs boson from these destabilizing tendencies (supersymmetry, CFT, extra dimension, etc.), or have banished the offending  fundamental scalar from nature (technicolor, compositeness, higgsless models, etc.), while other ideas have found ways to push the problem to higher scales (little Higgs theories, etc.).

Of course there are too many ideas out there for all to be correct. Nevertheless, if there are a thousand ideas and only one is right, it does  not mean that the others were useless, just as when a thousand rescue volunteers are looking for a little girl lost in the woods and one finds her, it does not mean the others were useless.  It may be argued that the only useless ideas are ones not grounded in rigor or are incompatible with past observations. This criteria for the worth of an idea is somewhat looser than the criteria we normally apply to theory in deciding what is good work.  Normally, we give our highest esteem to efforts that solve problems.   We value invention over unmoored creativity. I once heard an inventor describe what he does as first asking ``What sucks?" and then working day and night to make it better.
That is what we mostly do in physics.  We worship inventions. We dislike the SM Higgs boson and its quantum instability. This leads us to invent technicolor, supersymmetry, extra dimensions, etc.\  and then further invent solutions to their iatrogenic illnesses.  This formula is rather human-centric because we care most about {\it our} problems -- at the core, they are the problems associated with understanding the particles that make up our bodies. Surely, there is more to the universe than that.

A more universalist approach asks rather ``what's possible?"  There is great danger in this approach, since a whole lot more things are possible than are even probable. What then can discipline us? A new answer to this question is the Large Hadron Collider.  
The LHC era beckons us to approach physics less as an inventor and more humbly as a universalist. The beckoning is due to the filtering opportunity of experiment, and the impertinent susurrations that we shall fall short if we only take seriously our inventions.
Agreeing to the LHC as the primary disciplinarian of our creativity can yield a deeper interpretation of the data and perhaps may lead to new discoveries that were not anticipated. 

Thus, it is the existence of the LHC that propels  me to write about hidden worlds, or 
hidden sectors\cite{Schabinger:2005ei,Kumar:2006gm,Bowen:2007ia,Gopalakrishna:2008dv,Strassler,others}.  I could argue some second-order problem-solving explanation for why we must all care about this issue, by telling you that many ideas of  physics beyond the SM have sectors in addition to the SM that are hard to get rid of. I could also describe why  landscape studies imply the existence of even hundreds of possible new sectors\cite{Dijkstra:2004cc} that have nothing directly to do with solving any deep problem in nature that we recognize. No, instead, despite the motivating paralepsis, the physics of this chapter has but one core reason for cogitation: it can be discovered at the LHC. 

\section*{Hidden Worlds}

The definition of ``hidden" that I use here is the collection of particles that are not the SM, that are not charged under SM gauge groups, and that do not couple via gauge interactions to SM particles.  The possibilities are numerous. We can envision analogous copies of the SM charged under new gauge groups $SU(3)'_c\times SU(2)'_L\times U(1)'_Y$. We can envision pure singlet states. We can envision gauge fields of exotic gauge groups of large dimensionality. Very little experimental data bears on the question of whether such sectors exist.

It is not assured that we shall be able to discern the existence of a hidden world. All we can do is identify opportunities and explore them.   Of course, any gauge invariant and Lorenz invariant operator of the SM ${\cal O}^{inv}_{SM}$ can be paired with a similar operator from the hidden sector ${\cal O}^{inv}_{hid}$ to form ${\cal O}^{inv}_{SM}{\cal O}^{inv}_{hid}$.  If this resulting operator is irrelevant (dimension $> 4$) it will be suppressed by some unknown scale $M_*$.  We have no {\it a piori} idea what scale $M_*$ should be; however, we know that if it is above a few TeV it is unlikely we shall see evidence of this interaction due to decoupling.

The SM however does have two operators that are gauge-invariant and relevant (dimension $<4$):
the hypercharge field strength tensor $B_{\mu\nu}$ and the Higgs modulus squared $|H_{SM}|^2$.
These two operators give us hope that we can couple to a hidden sector in a  relevant or marginal
way (dimension $\leq 4$), thereby enabling a search for a  hidden world via the hypercharge gauge boson or the Higgs boson of the SM. 

Indeed, both of these operators can be exploited in the above-stated way to explore the simplest, non-trivial hidden sector that couples to $B_{\mu\nu}$ and $|\Phi_{SM}|^2$: $U(1)_X$ gauge theory with a complex Higgs boson $\Phi_H$ that breaks the symmetry upon condensation.  We call this simple model the ``Hidden Abelian Higgs Model" or HAHM, and explore the rich phenomenology that it implies for the LHC.

\section*{Hidden Abelian Higgs Model (HAHM)}

In this section I define precisely what I mean by HAHM.  First, 
we have the afore-mentioned extra $U(1)_X$ factor in addition to the SM gauge group.
The only coupling of this new gauge sector to the SM is through kinetic mixing with
the hypercharge gauge boson $B_\mu$.
The kinetic energy terms of the $U(1)_X$ gauge group are
\beq
{\cal L}^{KE}_X = -\frac{1}{4} \hat{X}_{\mu\nu} \hat{X}^{\mu\nu} + \frac{\chi}{2} \hat{X}_{\mu\nu} \hat{B}^{\mu\nu} \ ,
\eeq
where we comment later that $\chi \ll 1$ is helpful to keep precision electroweak predictions consistent with experimental measurements.

We introduce a new Higgs boson $\Phi_{H}$ in addition to the usual SM Higgs boson
$\Phi_{SM}$.
Under $SU(2)_L \otimes U(1)_Y \otimes U(1)_X$ we take the representations
$\Phi_{SM}: (2, 1/2, 0)$ and $\Phi_{H}: (1, 0, q_X)$, with $q_X$ arbitrary.
The Higgs sector Lagrangian is
\bea 
{\cal L}_{\Phi} &=& |D_\mu \Phi_{SM}|^2
+ |D_\mu \Phi_H |^2 
 + m^2_{\Phi_H}|\Phi_H|^2 + m^2_{\Phi_{SM}}|\Phi_{SM}|^2 \nonumber\\
& & - \lambda|\Phi_{SM}|^4 - \rho|\Phi_H|^4 - \kappa
|\Phi_{SM}|^2|\Phi_H|^2, \label{Lphi.EQ} 
\eea
so that $U(1)_X$ is broken spontaneously by $\left< \Phi_H \right> = \xi/\sqrt{2}$,
and electroweak symmetry is broken spontaneously as usual by
$\left< \Phi_{SM}\right> = (0,v/\sqrt{2})$. 

One can diagonalize the kinetic terms by redefining $\hat{X}_\mu ,
\hat{Y}_\mu \rightarrow X_\mu , Y_\mu$ with
\begin{displaymath}
\left( \begin{array}{c} X_\mu \\ Y_\mu \end{array} \right) = \left(
\begin{array}{cc} \sqrt{1-\chi^2} & 0 \\ -\chi & 1 \end{array} \right)
\left( \begin{array}{c} \hat{X}_\mu \\ \hat{Y}_\mu \end{array}
\right). \label{higgs.eq}
\end{displaymath}
The covariant derivative is then\beq D_\mu =
\partial_\mu + i (g_X Q_X + g^\prime \eta Q_Y) X_\mu + i g^\prime
Q_Y B_\mu + i g T^3 W^3_\mu \  . \label{DMUGAU.EQ} \eeq
where $\eta \equiv \chi / \sqrt{1-\chi^2}$.

After a $GL(2,R)$ rotation to diagonalize the kinetic terms followed by an $O(3)$ rotation
to diagonalize the $3 \times 3$ neutral gauge boson mass matrix, we can write the
mass eigenstates as (with $s_x\equiv \sin{\theta_x}$,  $c_x\equiv \cos{\theta_x}$)
\bea
\begin{pmatrix} B \\ W^3 \\ X  \end{pmatrix} =
\begin{pmatrix}
c_W & -s_W c_\alpha  & s_W s_\alpha \\
s_W & c_W c_\alpha & -c_W s_\alpha \\
0 & s_\alpha & c_\alpha
\end{pmatrix}
\begin{pmatrix} A \\ Z \\ Z'  \end{pmatrix} \ ,
\label{GAU2MAS.EQ}
\eea
where the
usual weak mixing angle and the new gauge boson mixing angle  are
\beq
s_W \equiv \frac{g^\prime}{\sqrt{g^2 + {g^\prime}^2}} \ ; \quad
\tan{\left( 2\theta_\alpha \right)} = \frac{-2 s_W \eta}{1 - s_W^2\eta^2 - \Delta_Z} \ ,
\label{tan2al.EQ}
\eeq
with $\Delta_Z =
M_{X}^2/M_{Z_0}^2$, $M_{X}^2 = \xi^2 g_X^2 q_X^2$, $M_{Z_0}^2 =
(g^2 + {g^\prime}^2) v^2 / 4$. $M_{Z_0}$ and $M_X$ are masses before mixing. The photon is massless (i.e., $M_A = 0 $), and the two heavier gauge boson
mass eigenvalues are 
\begin{eqnarray} M_{Z, Z^\prime} &=& \frac{M_{Z_0}^2}{2}
\left[\left(1+s_W^2 \eta^2 + \Delta_Z \right) \right. \nonumber\\ &&
\left. \qquad \pm \sqrt{\left( 1 - s_W^2 \eta^2 - \Delta_Z \right)^2
+ 4 s_W^2 \eta^2 } \right] ,
\end{eqnarray}
valid for $\Delta_Z < (1-s_W^2 \eta^2)$ ($Z \leftrightarrow Z^\prime$ otherwise). Since we assume that $\eta \ll 1$, mass
eigenvalues are taken as $M_Z \approx M_{Z_0}=91.19$ GeV and 
$M_{Z'} \approx M_X$.

The two real physical Higgs bosons $\phi_{SM}$ and $\phi_H$ mix after symmetry breaking,
and the mass eigenstates $h, H$ are
\begin{displaymath} \left( \begin{array}{c} \phi_{SM} \\ \phi_H
\end{array} \right) = \left( \begin{array}{cc} c_h & s_h \\ -s_h &
c_h \end{array} \right) \left( \begin{array}{c} h \\ H \end{array}
\right).
\end{displaymath}
Mixing angle and mass eigenvalues are 
\begin{eqnarray}
\tan{(2\theta_h)} &=& \frac{\kappa v \xi}{\rho \xi^2 - \lambda v^2}
\
\\ M_{h,H}^2 = \left( \lambda v^2 + \rho \xi^2 \right)
            &\mp& \sqrt{ (\lambda v^2 - \rho \xi^2)^2 + \kappa^2 v^2 \xi^2} \ .
\end{eqnarray}

In summary, the model has been completely specified above. The effect of HAHM on LHC phenomenology is to introduce two extra physical states $Z'$ and $H$.  $Z'$ is an extra gauge boson mass eigenstate that interacts with the SM fields because of gauge-invariant, renormalizable kinetic mixing with hypercharge, and $H$ is an extra Higgs boson that interacts with the SM fields  because of renormalizable modulus-squared mixing with the SM Higgs boson.

The Feynman rules are obtained from a straightforward expansion of the above lagrangian in terms of mass eigenstates.  Some of the Feynman rules most relevant for LHC studies are given 
below\cite{Gopalakrishna:2008dv}.

\underline{Fermion couplings}:
Couplings to SM fermions are
\bea
\bar \psi\psi Z &:& \frac{ig}{c_W} \left[ c_\alpha ( 1 - s_W t_\alpha \eta ) \right]
\left[  T^3_L - \frac{(1 - t_\alpha \eta/s_W)}{( 1 - s_W t_\alpha \eta)}  s_W^2 Q  \right]
\nonumber \\
\bar\psi\psi\zpri &:& \frac{-ig}{c_W} \left[c_\alpha  (t_\alpha + \eta s_W) \right] \left[ T^3_L - \frac{ (t_\alpha + \eta/s_W)}{ (t_\alpha + \eta s_W)} s_W^2 Q \right]
\label{GBFerm.EQ}
\eea
where  $Q = T^3_L + Q_Y$ and $t_\alpha \equiv s_\alpha / c_\alpha$.
The photon coupling is as in the SM and is not shifted.

\underline{Triple gauge boson couplings}:
We ${\cal R}$ being the coupling relative to the corresponding SM, one finds
${\cal R}_{A W^+ W^-} = 1$, ${\cal R}_{Z W^+ W^-} = c_\alpha$ and
${\cal R}_{\zpri W^+ W^-} = - s_\alpha$ (the last is compared to the SM $Z W^+ W^-$ coupling).
We will normally assume rather small kinetic mixing and so  to leading order 
we have $c_\alpha \approx 1$, $s_\alpha \ll 1$.

\underline{Higgs couplings}:
The Higgs couplings are
\beq
\label{hZpZ_coup.EQ}
\begin{split}
 h f f &:
-i c_h \frac{m_f}{v} \ , \qquad
h W W :
2i c_h \frac{M^2_W}{v} \ , \\
h Z Z &:
2i c_h \frac{M^2_{Z_0}}{v} (-c_\alpha + \eta s_W s_\alpha)^2 - 2i s_h
\frac{M^2_X}{\xi} s^2_\alpha \ , \\
h \zpri \zpri &:
2i c_h \frac{M^2_{Z_0}}{v} (s_\alpha + \eta s_W c_\alpha)^2 -
2i s_h \frac{M^2_X}{\xi} c^2_\alpha \ , \\
h \zpri Z &:
2i c_h \frac{M^2_{Z_0}}{v} (-c_\alpha + \eta s_W s_\alpha)
(s_\alpha + \eta s_w c_\alpha)  -2i s_h \frac{M^2_X}{\xi} s_\alpha c_\alpha \ .
\end{split}
\eeq

\section*{Precision Electroweak}

Generally speaking HAHM does not have significant disruptions of the precision electroweak predictions compared to the SM to cause undo worry.  In other words, a vast region of parameter space is completely compatible with the precision electroweak data. However, it is useful to review some of the issues\cite{Bowen:2007ia,Kumar:2006gm}.

First, when the  $X$ gauge boson mixes with hypercharge there will be a shift in the precision electroweak observables compared to the SM. For example, from hypercharge-$X$ mixing, the $Z$ mass eigenvalue is further shifted relative to the $W^\pm$ mass.  These effects can be computing in an effective Peskin-Takeuchi parameter analysis\cite{Peskin:1991sw,Holdom:1990xp,Babu:1997st}. One finds that the three most important observables for constraining new physics
by precision electroweak observables are
\bea
\Delta m_W & =  & (17\mev)\, \Upsilon \\
\Delta \Gamma_{l^+l^-} &  =  & -(80\kev)\, \Upsilon   \\
\Delta \sin^2\theta_W^{eff} &  = &  -(0.00033)\,\Upsilon
\eea
where
\beq
\Upsilon\equiv \left(\frac{\eta}{0.1}\right)^2\left(\frac{250\gev}{m_X}\right)^2.
\eeq
Experimental measurements\cite{LEPEWWG}
imply that  $|\Upsilon|\lsim 1$. Kinetic mixing at the level of
$\eta \lsim {\cal O}(0.1)$ is not constrained if $M_X$ is greater than a few hundred GeV, and there is essentially no constraint  if $M_X$ is greater than about a TeV. This is consistent with the PDG analysis of constraints on other $Z'$ bosons\cite{PDG}.
For lighter $M_X$, which we will also concern ourselves with, the constraint is not difficult to satisfy as long as $\eta \lsim {\cal O}(0.01)$.

A pure singlet Higgs boson  causes no concern for  precision electroweak observables, but after mixing the  coupling of the Higgs to the gauge bosons is shared by two states of different masses. The leading order way to account for this is to first recognize that in the SM the Higgs boson mass constraints is succinctly summarized as\cite{LEPEWWG}
\beq
\log\left( M_{\rm Higgs}/1~{\rm GeV} \right) = 1.93^{+0.16}_{-0.17}.
\eeq
When two states, such as ours, mix and share the electroweak coupling, this constraint becomes to leading order
\beq
c_h^2 \log\left( \frac{M_h}{1~{\rm GeV}} \right) +  s_h^2 \log\left( \frac{M_H}{1~{\rm GeV}} \right)
\simeq 1.93^{+0.16}_{-0.17} \ .
\label{mhmH_fit.EQ}
\eeq
There is very little difficulty in exploring large regions of parameter space where the precision electroweak implications of this multi-Higgs boson theory are in agreement with all 
data\cite{Bowen:2007ia}.

Other constraints, such as perturbative unitarity and vacuum stability have been analyzed 
elsewhere\cite{Bowen:2007ia} and also can be accomodated easily within the theory.

\section*{Example LHC Phenomena of HAHM}

How do we find evidence for HAHM at colliders?  The main implication of HAHM is the different spectrum it implies compared to the SM spectrum of states:
\begin{itemize}
\item The existence of a new gauge boson $Z'$ that couples to SM states according to the strength of the kinetic mixing parameter.
\item The existence of two CP-even Higgs boson mass eigenstates, both of which couple to SM states by virtue of the mixing of the HAHM Higgs boson with the SM Higgs boson. 
\end{itemize}
These two simple qualitative facts, combined with the details of the HAHM langrangian enable us to explore many possible interesting implications for the LHC.

In the next few paragraphs I shall discuss a few of these implications. The reader should keep in mind that not all cases are simultaneously allowed by the theory. Each phenomenological manifestation I discuss can be considered the dominant interesting signal in a subset of the parameter space, not in all of parameter space.

\bigskip

\noindent
{\it Signal \#1: Universal suppression of Higgs boson signal}

\bigskip 

Let us suppose that the two Higgs bosons mix, such that the lightest Higgs boson is mostly SM, and the heavier Higgs boson eigenstate is mostly singlet.  Let us further suppose that the additional $Z'$ Higgs boson is sufficiently heavy or weakly coupled that is has no role in the phenomenology. In this case, the primary signal will be that the light Higgs mass eigenstate couples to the SM states in exactly the same way as the SM Higgs except there is a universal suppression of all interactions due to the mixing angle.  

Thus the cross-section is reduced by a factor of
\beq
\sigma(VV\to h)(m_{h})=c_h^2\sigma(VV\to h_{SM})(m_h)
\eeq
This implies that no state in the spectrum of Higgs bosons has a production cross-section as large as the SM Higgs boson, making production, and thus detection, more difficult.

Production is only half of the story when discussing detectability. One must also consider how the branching fraction changes. Of course, if there is only a universal suppression of couplings, the branching fractions will be identical to those of the SM Higgs boson.  However, if there are exotic matter states in the HAHM model in addition to just the $X$ boson and its associated symmetry-breaking $\Phi_H$ boson, the lightest Higgs might decay into them. If the exotic states are stable on detector time scales it would contribute to the invisible width of the Higgs boson $\Gamma^{\rm hid}$, which depends on exotic sector couplings, $m_h$ and $s_h^2$.  
The branching fraction into visible states is then reduced and computed by
\beq
B_i(h)=\frac{c_h^2\Gamma^{SM}_i(m_h)}{c_h^2\Gamma^{SM}_i(m_h)+\Gamma^{hid}(m_h)}.
\eeq

The effect of this universal suppression was studied in the context of hidden sectors\cite{Schabinger:2005ei} and also in a related context of extra-dimensional theories\cite{Wells:2002gq}. Of course, this signal is not unique to the HAHM, as any singlet Higgs boson that gets a vacuum expectation value could mimic it. However, a singlet with a vacuum expectation value is likely to have gauge charge, but it is not necessary that it be exclusively abelian and kinetic mix with hypercharge.  Thus, the universal suppression of the Higgs boson phenomenology is more general than just HAHM. 

\bigskip

\noindent
{\it Signal \#2: $H\to hh$}

\bigskip

Another broad implication of mixing with a singlet Higgs boson is the existence of a heavier Higgs boson that couples to SM states and can decay into a pair of lighter Higgs bosons.  This has been discussed in detail in the context of HAHM\cite{Bowen:2007ia}.  There, an example model was studied where 
\bea
m_H=300\gev,~~~m_h=115\gev,~~~c_h^2=\frac{1}{2}
\eeq
Thus, the decay of $H\to hh$ is kinematically allowed in this case, and the relevant branching fraction is $B(H\to hh)=1/3$.  

\begin{figure}[t]
	\centering
\includegraphics[angle=90]{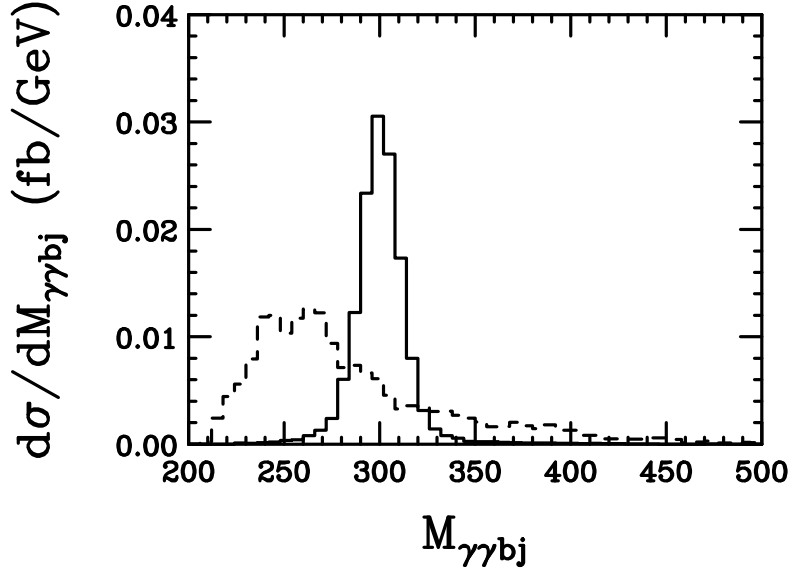}  
	\caption{Differential cross-section\cite{Bowen:2007ia} as a function of $b\bar b\gamma\gamma$ invariant mass in $H\to hh\to b\bar b\gamma\gamma$ production.}
	\label{fig:bbgammagamma}
\end{figure}

One of the most useful signals to find this decay chain is when one light Higgs decays to $h\to b\bar b$, which it is most apt to do, and the other decays to the rarer $h\to \gamma\gamma$.  The signal is reduced substantially by requiring this lower probability $b\bar b\gamma\gamma$ final state, but the 
background is reduced by even more.  It is found that with $30\xfb^{-1}$ the total expected signal event rate after relevant cuts and identification criteria are applied\cite{Bowen:2007ia} is 8.2 on a background of 0.3. 
Fig.~\ref{fig:bbgammagamma} shows the differential cross-section as a function of invariant mass of $b\bar b\gamma\gamma$ for these events.

In the above example the lighter higgs boson is light -- right at the edge of the current SM limits -- and decays preferentially  to $b\bar b$. If the lighter Higgs boson is heavier than this, the decay to $WW$ starts to become dominant. The cross-over point where $B(b\bar b)=B(W^+W^-)$ is about $m_h=130\gev$.  For this case of $m_h\gsim 130\gev$, it is more fruitful to exploit the $4W\to 4\ell +$ missing $E_T$ signature.  An analysis of this final state has been shown\cite{Bowen:2007ia} to be a promising approach to finding $H\to hh$ at the LHC.

\bigskip

\noindent
{\it Signal \#3: Trans-TeV Narrow Higgs Boson}

\bigskip

Within the SM the Higgs boson becomes so broad when its mass is above about $700\gev$ that it starts to become meaningless to even call it a particle. There is no sense in which a trans-TeV Higgs boson resonance can be found within standard Higgs boson phenomenology at the LHC. However, in the mixed boson sector induced by HAHM, we find that a Higgs boson just like the SM can exist, except its couplings are universally suppressed by a factor of $s_h^2$ compared to the SM Higgs boson. Thus, a reasonably narrow trans-TeV Higgs boson can be in the spectrum, and can be searched for at the LHC.

\begin{figure}[t]
	\centering
		\includegraphics[angle=90]{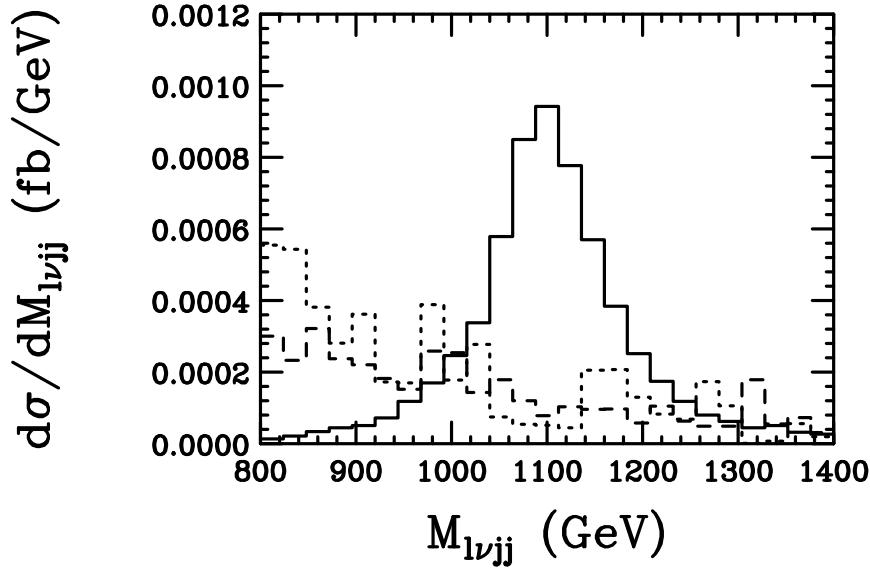}  
	\caption{Differential cross-section\cite{Bowen:2007ia} as a function of invariant mass of $\ell\nu_T jj$ for the signal $H\to WW\to \ell\nu jj$ (solid) and two main backgrounds,   $WWjj$ (dashed) and $t\bar t jj$ (dotted). }
	\label{fig:mlvjj}
\end{figure}

The narrowness of the Higgs boson is also correlated with a low production cross-section, and so the biggest challenge is simply getting enough events to even analyse.  Once they are produced, distinguishing them from background is made possible by the very high energy invariant mass and transverse mass reconstructions.  For example, in Fig.~\ref{fig:mlvjj} the invariant mass distribution of $l\nu jj$ (missing $E_T$ vector used for $\nu$) is plotted for the signal ($H\to WW\to l\nu jj$) of a $m_H=1.1\tev$ Higgs boson and compared to the distributions of the most significant backgrounds from $WWjj$ and $t\bar t jj$.  The cuts we applied were
\bea
& & p_T(e,\mu)>100\gev~{\rm and}~|\eta(e,\mu)|<2.0 \nonumber \\
&& {\rm Missing~}E_T>100\gev \nonumber \\
&& p_T(j,j)>100\gev~{\rm and}~m_{jj}=m_W\pm 20\gev \nonumber \\
&& {\rm ``Tagging~jets"~with~}|\eta|>2.0\nonumber
\eea
With $100\xfb^{-1}$ the signal gives 13 events in the invariant mass range between 1.0 and 1.3 TeV, compared to a background of 7.7 events. This is obviously not ``early phase" LHC signal, and it highlights the challenges in finding evidence of heavy Higgs bosons from a hidden sector. Nevertheless, it is possible to find evidence for it with enough integrated luminosity, which the LHC should attain in time.
The signal significance will increase when all possible channels are included.

\begin{figure}[t]
	\centering
\includegraphics[angle=90]{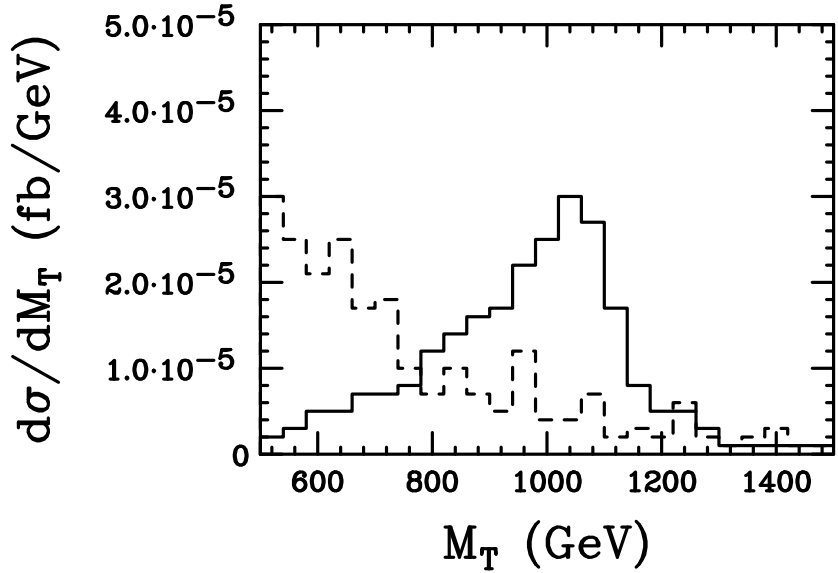}  
	\caption{Differential cross-section\cite{Bowen:2007ia} of transverse mass of $Z$ and missing $E_T$ for the signal $H\to \ell\ell \nu\nu$ (solid) and the main background $ZZjj$ (dashed).}
	\label{fig:mT}
\end{figure}

An even more challenging final state to consider is $H\to ZZ\to ll\nu\nu$. The most significant background is $ZZjj$. The cuts we applied were
\bea
&& p_T(l^+,l^-)>100\gev~{\rm and}~|\eta(l^+,l^-)|<2.0 \nonumber \\
&& m_{ll}=m_Z\pm 5\gev \nonumber \\
&& {\rm Missing~}E_T>100\gev \nonumber \\
&& {\rm ``Tagging~jets"~with~}|\eta|>2.0\nonumber
\eea
Fig.~\ref{fig:mT} shows  the transverse mass of $Z$ with missing $E_T$ for a different signal topology ($H\to ZZ\to \ell\ell \nu\nu$) and compared with the most significant background, $ZZjj$. 
If we restrict ourselves to $0.8\tev<M_T<1.4\tev$ with $500\xfb^{-1}$ there are 3.9 signal events compared to 1.4 background events. Again, this is not early stage LHC physics. Finding and studying this kind of trans-TeV Higgs boson physics should be considered a strong motivation for the high-luminosity phase of the LHC.

\bigskip

\noindent
{\it Signal \#4: Searching for $Z'$ resonance}

\bigskip

When the exotic $X$ boson mixes with hypercharge via kinetic mixing, the resulting mass eigenstates picks up couplings with the SM states.  At the LHC one can look for resonance production and decay of this new $Z'$ boson. One of the best approaches expeimentally to finding evidence for such a $Z'$ is to investigate the $\mu^+\mu^-$ invariant mass spectrum.  

\begin{figure}[t]
	\centering
\includegraphics[scale=0.55]{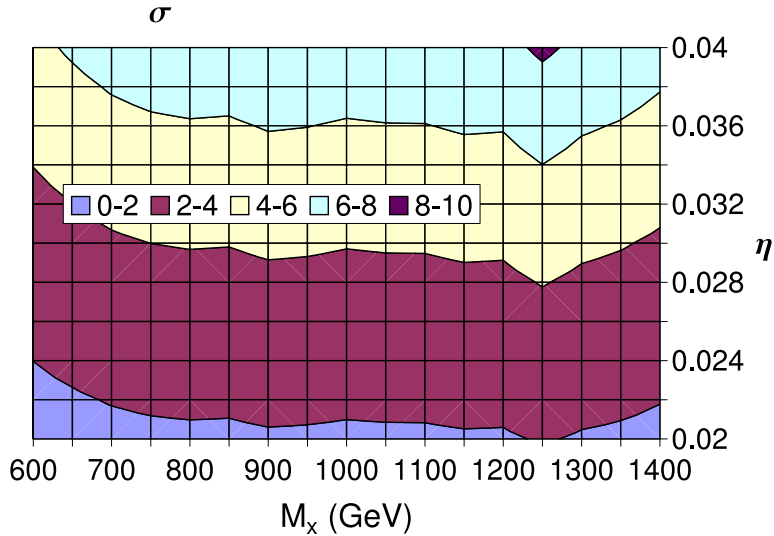}  
	\caption{LHC detection prospects\cite{Kumar:2006gm} in $\eta-M_X$ plane for $100\xfb^{-1}$ integrated luminosity at the LHC. Signal significance exceeds 5 only when $\eta\gsim 0.03$.}
	\label{fig:etaMX}
\end{figure}

There is a staggeringly large literature on the search for $Z'$ bosons at colliders\cite{ZprimeStudies}, but usually little emphasis is put on treating the overall coupling strength as a free parameter that could be very 
small\cite{Chang:2006fp}. Indeed, the kinetic mixing is normally expected to be maybe loop level for theories of this kind, which would imply a rather small coupling of the $Z'$ to SM states.  We studied some of the implications of very weakly coupled $Z'$ physics for the LHC\cite{Kumar:2006gm}. The summary graph of this study is Fig.~\ref{fig:etaMX} where it was determined that it is very hard for the LHC to probe lower than $\eta \sim 10^{-2}$, which is not particularly constraining to the theory given expectations.

\bigskip

\noindent
{\it Signal \#5: $h\to Z'Z'\to 4l$}

\bigskip

In the previous discussion we noted that values of $\eta\lsim 10^{-2}$ are not very well constrained by the data, nor will they be easily constrained at the LHC.  Nevertheless, even a tiny value of $\eta$
has important phenomenological implications to collider physics. Since a tiny value of $\eta\lsim 10^{-2}$ may even be more motivated, it becomes interesting to ask what effect it could still have on LHC phenomenology. In a recent paper\cite{Gopalakrishna:2008dv}, we showed  that a light $Z'$ boson with $\eta\sim 10^{-4}$ could lead to large branching fractions of $h\to Z'Z'\to 4f$, where the first step $h\to Z'Z'$ is accomplished by Higgs mixing and a sufficiently light $Z'$ mass, and the last step $Z'\to f\bar f$ is allowed merely because $\eta\neq 0$ and the $Z'$ must decay.

\begin{figure}[t]
	\centering
\includegraphics[scale=0.7]{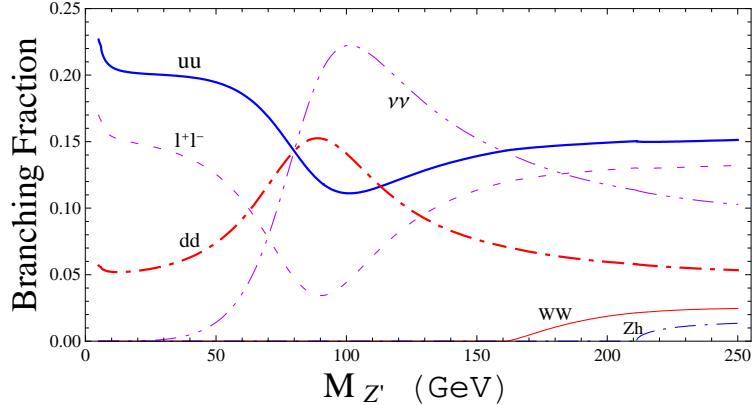}  
	\caption{The branching ratio\cite{Gopalakrishna:2008dv} of the $Z'$ boson as a function of its mass for the parameters $c_h=0.5$ and $\eta=10^{-4}$.}
	\label{fig:xdecay}
\end{figure}

The branching fractions of the $Z'$ depend on several factors in the theory, but to illustrate they are shown in Fig.~\ref{fig:xdecay} for $c_h^2=0.5$ and $\eta=10^{-4}$. The branching fraction into four leptons is high enough to exploit its clean signatures at the LHC. Looking for various invariant mass peaks and making various kinematic cuts on the data, the prospects of finding this signature at the LHC with only a few $\xfb^{-1}$ are 
excellent\cite{Gopalakrishna:2008dv}
provided the two Higgs bosons mix significantly and $h\to Z'Z'$ is kinematically accessible. This may even be the channel where the light Higgs boson mass is first discovered, since it is an easy ``gold-plated channel". Compare that with the very difficult normal searches of the Higgs boson with mass $\sim 120\gev$, which is made even more difficult when its production cross-section is reduced, by $50\%$ in this case.

\section*{Beyond the Standard Model and the Hidden World}

The discussion in this chapter has all been about physics that attaches itself to the SM relevant operators. However, there are many reasons to believe that the SM sector cannot stand alone in the operation of symmetry breaking and mass generation. We expect new non-hidden particles, such as superpartners and KK excitations, etc., that solve the problems we have identified with the SM.   
Some people wish that we will discover something totally new and unexpected. However, it would necessitate a  shift in our philosophical approach to frontier basic science.

The LHC is just as much a philosophy experiment as a physics experiment.   The impacting issue is ``To what degree can humans discern nature from pure thought?"  Arguably we have had some success already in the past, but would anything in the past compare, for example, to postulating that  supersymmetry cures the hierarchy problem if it turns out to be correct?  It would certify that humans can see around the corner and discern deep new principles into the energy frontier.  If we get that right, no idea would be too esoteric, and no scale would be too remote or inaccessible for humans to discuss with confidence and expectation for understanding.  

Thus, I hope and expect that we find new physics that explains by principle the stability of the electroweak scale from ideas that we have already developed. How does this impact the HAHM discussion presented above? First, if it is supersymmetry then it is likely to merely complicate the discussion above, as many new states will be produced and will decay in the detector, and the number of Higgs bosons will be  greater, making simple mixing angle factors such as $c_h$ from our $2\times 2$ matrix 
into more complicated combinations of mixing angles such as $c_{\theta_1}c_{\theta_2}s_{\theta_3}$. The origin of the ``hidden sector" higgs mixing with MSSM Higgs is most likely to be from the  renormalizable coupling in the superpotential: $SH_uH_d$, which yields $|S|^2|H_i|^2$-type couplings in the $F$-term lagrangian directly analogous to the $|\Phi_H|^2|\Phi_{SM}|^2$ mixing terms we have discussed here. Thus, the basic ideas shine through in the Higgs sector and analyses similar to those discussed above can be applicable.

Of course, if the stability of the electroweak scale is explained by the banishment of all fundamental scalars from nature, then additional Higgs boson mixing may not be relevant, but perhaps an effective Higgs boson mixing theory with composite Higgs bosons would be a useful description. This would be highly model dependent, and the data from LHC will have to guide us to decide if there is a path by which we can interpret electroweak symmetry breaking by effective Higgs boson scalars.
If so, looking for a Hidden World then would be possible again via couplings to this effective Higgs boson.

\bigskip

\noindent
{\it Acknowledgments:} I would like to thank my collaborators for the many interesting discussions on these topics: M. Bowen, Y. Cui, S. Gopalakrishna, S. Jung, J. Kumar, and R. Schabinger. This work is supported in part by CERN, the Department of Energy, and the Michigan Center for Theoretical Physics (MCTP).


\end{document}